\documentclass[aps,showpacs,preprintnumbers,amsmath, amssymb]{revtex4}

\usepackage{float}
\usepackage{graphics,epsfig}
\usepackage{graphicx}
\usepackage{dcolumn}
\usepackage{bm}

\begin{document}
\baselineskip=0.8 cm
\title{{\bf General holographic superconductor models with backreactions}}

\author{Qiyuan Pan$^{1,2}$, Bin Wang$^{1}$}
\affiliation{$^{1}$ INPAC and Department of Physics, Shanghai Jiao Tong University, Shanghai 200240, China}
\affiliation{$^{2}$ Institute of
Physics and Department of Physics, Hunan Normal University, Changsha, Hunan 410081, China}

\vspace*{0.2cm}
\begin{abstract}
\baselineskip=0.6 cm
\begin{center}
{\bf Abstract}
\end{center}

We study general models of holographic superconductors away from the probe limit.  We find that the backreaction of the spacetime
can bring richer physics in  the phase transition. Moreover we observe that the ratio $\omega_{g}/T_{c}$ changes with the strength
of the backreaction and is not a universal constant.

\end{abstract}

\maketitle
\newpage
\vspace*{0.2cm}

\section{Introduction}

The anti-de Sitter/conformal field theories (AdS/CFT) correspondence
\cite{Maldacena,Witten,Gubser1998} states that a $d$-dimensional
strongly coupled conformal field theory on the boundary is
equivalent to a $(d+1)$-dimensional weakly coupled dual
gravitational description in the bulk. This remarkable finding
provides a novel method to study the strongly coupled system at
finite density and builds a useful connection between the condensed
matter and the gravitational physics (for reviews, see Refs.
\cite{HartnollRev,HerzogRev,HorowitzRev}). It has been shown that
the spontaneous $U(1)$ symmetry breaking by bulk black holes can be
used to construct gravitational duals of the transition from normal
state to superconducting state in the boundary theory
\cite{GubserPRD78}. In \cite{HartnollPRL101} the $(2+1)$-dimensional
superconductor which consists of a system with a black hole and a
charged scalar field in 3+1 dimensions was introduced in the probe
limit where the backreaction of the matter fields on the metric is
small and can be neglected. It is interesting to note that the
properties of a $(2+1)$-dimensional superconductor can indeed be
reproduced in this simple model. Motivated by the application of the
Mermin-Wagner theorem to the holographic superconductors, there have
been investigations on the effects of the curvature corrections on
the ($3+1$)-dimensional superconductor and higher dimensional ones.
It was shown that the Gauss-Bonnet coupling affects the condensation
of the scalar field \cite{Gregory,Pan-Wang,Ge-Wang,Liu-Wang} and the
vector field \cite{Cai-Nie-Zhang}, and the higher curvature
correction makes the condensation harder to form. Further the
curvature correction term will cause the unstable ratio
$\omega_{g}/T_c$ \cite{Gregory,Pan-Wang}. The gravity models with
the property of the so-called holographic superconductor have
attracted considerable interest for their potential applications to
the condensed matter physics, see for example
\cite{HorowitzPRD78,Amado,Koutsoumbas,Maeda79,Sonner,
Cai-Zhang,Jing-Chen,Nishioka,Siopsis,Chen-Wen,Gao-Zhang,Kuang-Ling}.

A lot of studies in the holographic superconductor disclosed a
second order phase transition in strongly interacting systems using
the AdS/CFT correspondence. It was observed that a fairly wide class
of phase transitions can be allowed if one generalizes the basic
holographic superconductor model in which the spontaneous breaking
of a global U(1) symmetry occurs via the St\"{u}ckelberg mechanism
\cite{FrancoJHEP}. This framework allows tuning the order of the
phase transition which can accommodate the first order phase
transition to occur, and for the second order phase transition it
allows tuning the values of critical exponents \cite{FrancoPRD}. An
interesting extension was done in \cite{Aprile-Russo} by
constructing general models for holographic superconductivity. It
was found that except some universal model independent features,
some important aspects of the quantum critical behavior strongly
depend on the choice of couplings, such as the order of the phase
transition and critical exponents of second-order phase transition.
In addition to the numerical investigation, analytical understanding
on the phase transition of holographic superconductor was also
provided in \cite{Herzog-2010}. Rich phenomena in the phase
transition were also found for the holographic superconductors in
Einstein-Gauss-Bonnet gravity where the Gauss-Bonnet coupling can
play the role in determining the order of phase transition and
critical exponents in the second-order phase transition
\cite{Pan-WangPLB}.

Most of avaliable studies on the holographic superconductors are
limited in the probe approximation, although they can give most
qualitative results on the holographic superconductivity, the study
on the effect of backreaction is called for.  Recently there have
been a lot of interest to study the holographic superconductor away
from the probe limit and take the backreaction of the spacetime into
account\cite{HartnollJHEP12,Aprile-Russo,Liu-Sun,HorowitzJHEP11,Gubser-Nellore,
Ammon-et,Brihaye,Barclay-Gregory,Siani,GregoryRev,CaiBackreaction}.
Considering the backreaction, it was found that even the uncharged
scalar field can form a condensate in the $(2+1)$-dimensional
holographic superconductor model \cite{HartnollJHEP12}. In the
p-wave superfluids system, it was argued that the order of the phase
transition depends on the backreaction, i.e., the phase transition
that leads to the formation of vector hair changes from the second
order to the first order when the gravitational coupling is large
enough \cite{Ammon-et}. However this result was not observed in
studying the backreaction of (3+1)-dimensional holographic
superconductor in Einstein Gauss-Bonnet gravity \cite{Brihaye}.

It would be of great interest to further explore the holographic superconductivity with backreactions. In this work we will consider the effect
of the spacetime backreaction on the sufficiently general gravity dual describing a system of a U(1) gauge field and the scalar field coupled via
a generalized St\"{u}ckelberg Lagrangian. We will examine the effects of the backreaction on the condensation of the scalar hair and
conductivity. Furthermore we will investigate the phase transition when taking the backreaction into account. We will also generalize the
discussion to the Einstein-Gauss-Bonnet gravity by considering the combined effects of the generalized St\"{u}ckelberg mechanism and the
backreaction.

\section{$(2+1)$-dimensional superconducting models with backreaction}

We study the formation of scalar hair on the background of AdS black
hole in (3+1)-dimensions. The generalized action containing a U(1)
gauge field and the scalar field coupled via a generalized
St\"{u}ckelberg Lagrangian reads
\begin{eqnarray}\label{System}
S= \int d^{4}x\sqrt{-g}\left[\frac{1}{16\pi
G_{4}}\left(R-2\Lambda\right)+\mathcal{L}_{matter} \right],
\end{eqnarray}
where $G_{4}$ is the 4-dimensional Newton constant and
$\Lambda=-3/L^{2}$ is the cosmological constant.
$\mathcal{L}_{matter}$ is the generalized St\"{u}ckelberg Lagrangian
\cite{FrancoJHEP}
\begin{eqnarray}\label{ModelL}
\mathcal{L}_{matter}=-\frac{1}{4}F_{\mu\nu}F^{\mu\nu}-\frac{1}{2}\partial_{\mu}\psi\partial^{\mu}\psi
-\frac{1}{2}m^2\psi^2-\frac{1}{2}|\mathfrak{F}(\psi)|(\partial_{\mu}p-A_{\mu})
(\partial^{\mu}p-A^{\mu}),
\end{eqnarray}
where $\mathfrak{F}(\psi)$ is a general function of $\psi$. Here we
consider the simple form of $\mathfrak{F}(\psi)$
\begin{eqnarray}\label{ModelF}
\mathfrak{F}(\psi)=\psi^{2}+c_{4}\psi^{4},
\end{eqnarray}
which has been discussed in \cite{Gubser-C-S-T,Gauntlett-Sonner}.
Our study can be easily extended to a more general form
$\mathfrak{F}(\psi)=\psi^{2}+c_{\lambda}\psi^{\lambda}+c_{4}\psi^{4}$
with the model parameters $c_{\lambda}$, $c_{4}$ and
$\lambda\in[3,4]$ just as discussed in \cite{FrancoPRD,Pan-WangPLB},
which will not alter the qualitative result.  Considering the gauge
symmetry
\begin{eqnarray}
A_{\mu}\rightarrow
A_{\mu}+\partial_{\mu}\tilde{\Lambda},~~p\rightarrow
p+\tilde{\Lambda},
\end{eqnarray}
we can fix the gauge $p=0$ by using the gauge freedom.

We are interested in including the backreaction so we use the ansatz
of the geometry of the 4-dimensional AdS black hole with the form
\begin{eqnarray}\label{BH metric}
ds^2=-g(r)e^{-\chi(r)}dt^{2}+\frac{dr^2}{g(r)}+r^{2}(dx^{2}+dy^{2}),
\end{eqnarray}
whose Hawking temperature, which will be interpreted as the temperature of the CFT, reads
\begin{eqnarray}\label{Hawking temperature}
T_{H}=\frac{g^{\prime}(r_{+})e^{-\chi(r_{+})/2}}{4\pi}.
\end{eqnarray}
$r_{+}$ is the black hole horizon defined by $g(r_{+})=0$. Choosing
the electromagnetic field and the scalar field as
\begin{eqnarray}
A=\phi(r)dt,~~\psi=\psi(r),
\end{eqnarray}
we can obtain the equations of motion
\begin{eqnarray}\label{Psi-Phi}
&&\chi^{\prime}+\gamma\left[\frac{r}{2}\psi^{\prime
2}+\frac{r}{2g^{2}}e^{\chi}\phi^{2}\mathfrak{F}(\psi)\right]=0,\nonumber\\
&&g^{\prime}-\left(\frac{3r}{L^{2}}-\frac{g}{r}\right)+\gamma
rg\left[\frac{1}{4}\psi^{\prime 2}+\frac{1}{4g}e^{\chi}\phi^{\prime
2}+\frac{m^{2}}{4g}\psi^{2}+\frac{1}{4g^{2}}e^{\chi}\phi^{2}\mathfrak{F}(\psi)\right]=0,
\nonumber\\ &&
\phi^{\prime\prime}+\left(\frac{2}{r}+\frac{\chi^{\prime}}{2}\right)\phi^\prime-\frac{\mathfrak{F}(\psi)}{g}\phi=0,\nonumber\\
&& \psi^{\prime\prime}+\left(\frac{2}{r}-\frac{\chi^{\prime}}{2}+
\frac{g^\prime}{g}\right)\psi^\prime
-\frac{m^2}{g}\psi+\frac{1}{2g^2}e^{\chi}\phi^2\mathfrak{F}^\prime(\psi)=0,
\end{eqnarray}
where the parameter $\gamma=16\pi G_{4}$. In the probe limit where
$\gamma\rightarrow0$, (\ref{Psi-Phi}) goes back to the
(2+1)-dimensional holographic superconductor model studied in
\cite{FrancoPRD}. For nonzero $\gamma$ we take the backreaction of
the spacetime into account.

The analytic solutions to Eq. (\ref{Psi-Phi}) for $\psi(r)=0$ are
the asymptotically AdS Reissner-Nordstr\"{o}m black holes
\begin{eqnarray}
g=\frac{r^{2}}{L^{2}}-\frac{2M}{r}+\frac{\gamma\rho^{2}}{4r^{2}}\,,\hspace{0.5cm}
\phi=\rho\left(\frac{1}{r_{+}}-\frac{1}{r}\right)\,,\hspace{0.5cm}\chi=0,
\end{eqnarray}
where $M$ and $\rho$ are the integration constants that can be
interpreted as the mass and the charge density of the solution,
respectively. When $\gamma=0$, the metric coefficient $g$ goes back
to the Schwarzschild AdS black hole. In order to get the solutions
with nonzero $\psi(r)$, we have to count on the numerical method
which has been explained in detail in
\cite{Aprile-Russo,Liu-Sun,HartnollJHEP12}. The equations of motion
(\ref{Psi-Phi}) can be solved numerically by doing integration from
the horizon out to the infinity. At the asymptotic AdS boundary
($r\rightarrow\infty$), the scalar and Maxwell fields behave like
\begin{eqnarray}
\psi=\frac{\psi_{-}}{r^{\lambda_{-}}}+\frac{\psi_{+}}{r^{\lambda_{+}}}\,,\hspace{0.5cm}
\phi=\mu-\frac{\rho}{r}\,, \label{infinity}
\end{eqnarray}
with
\begin{eqnarray}
\lambda_\pm=\frac{1}{2}(3\pm\sqrt{9+4m^{2}}~),
\end{eqnarray}
where $\mu$ and $\rho$ are interpreted as the chemical potential and
charge density in the dual field theory, respectively. Notice that
both of the falloffs are normalizable for $\psi$, so one can impose
boundary condition that either $\psi_{+}$ or $\psi_{-}$ vanishes
\cite{HartnollPRL101,HartnollJHEP12}. For simplicity, we will take
$\psi_{-}=0$ and the scalar condensate is now described by the
operator $\langle{\cal O}_{+}\rangle=\psi_{+}$. In this work we will
discuss the condensate $\langle{\cal O}_{+}\rangle$ for fixed charge
density. Moreover, we will consider the values of $m^{2}$ which must
satisfy the Breitenlohner-Freedman (BF) bound $m^{2}\geq-(d-1)^2/4$
\cite{BF} for the dimensionality of the spacetime $d=4$ in the
following analysis.

\subsection{The effects on the
scalar condensation and phase transition}

\begin{figure}[H]
\includegraphics[scale=0.75]{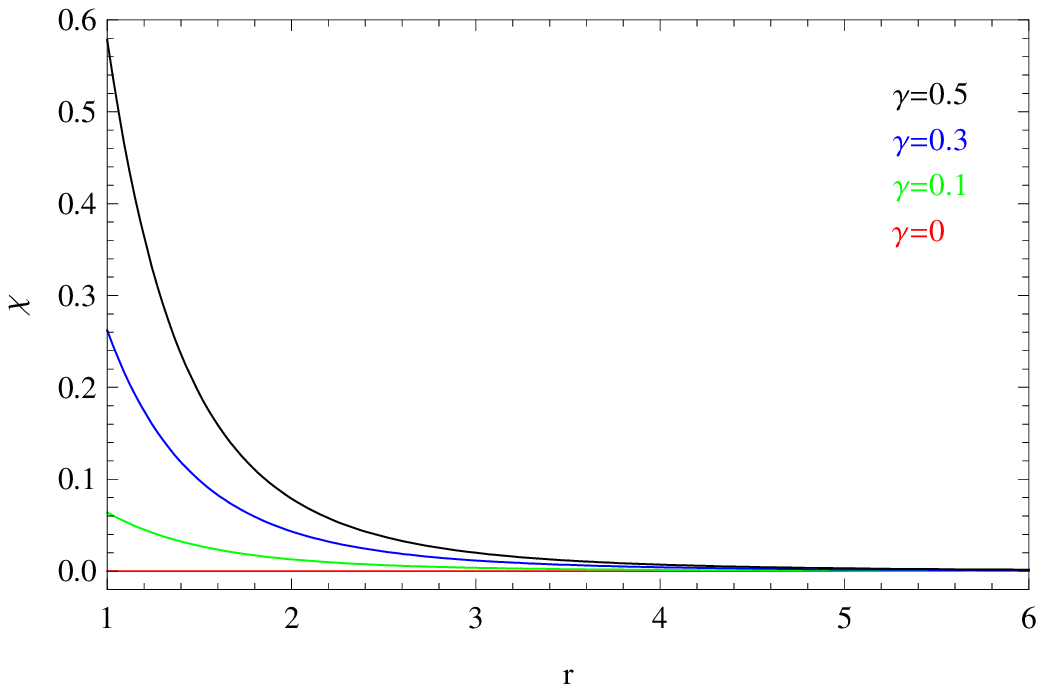}\hspace{0.2cm}
\includegraphics[scale=0.75]{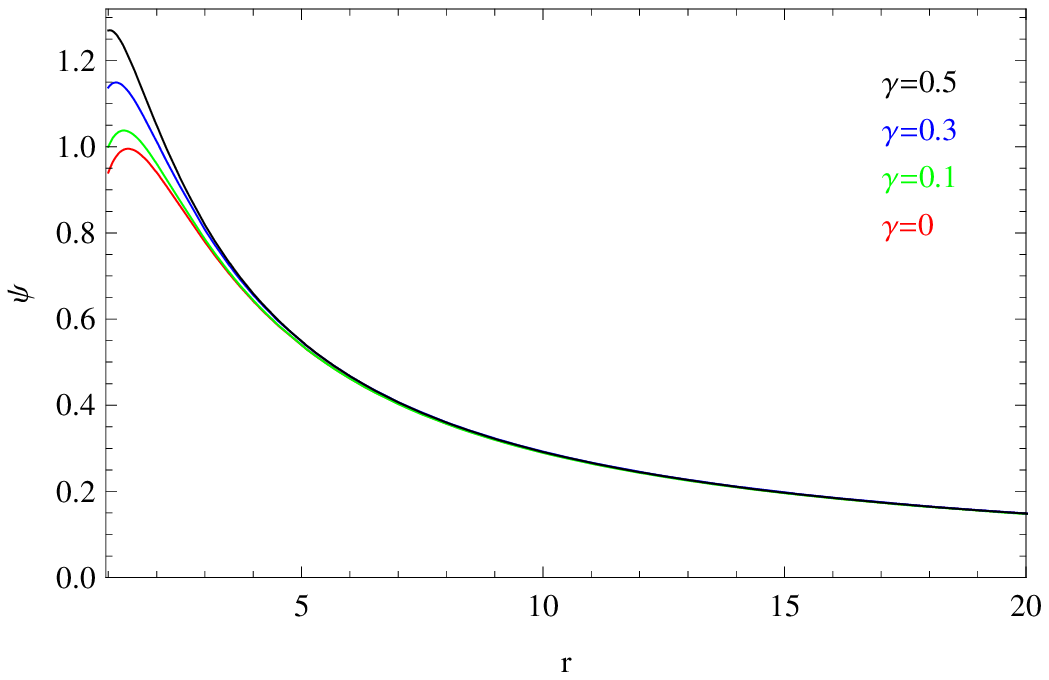}\\ \vspace{0.0cm}
\includegraphics[scale=0.75]{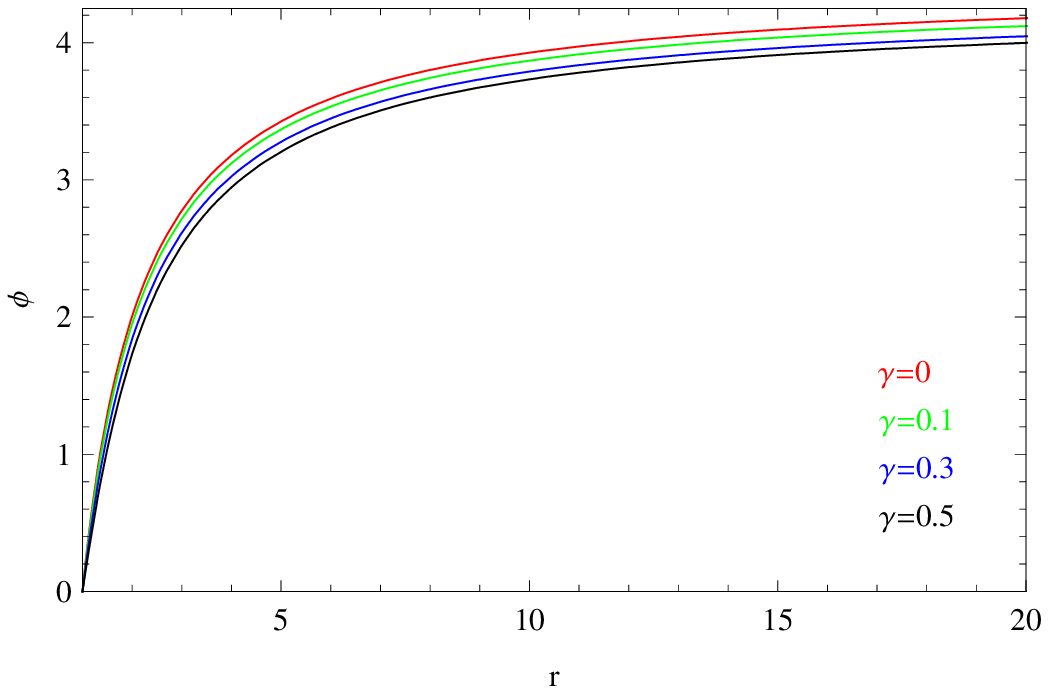}\hspace{0.2cm}
\includegraphics[scale=0.75]{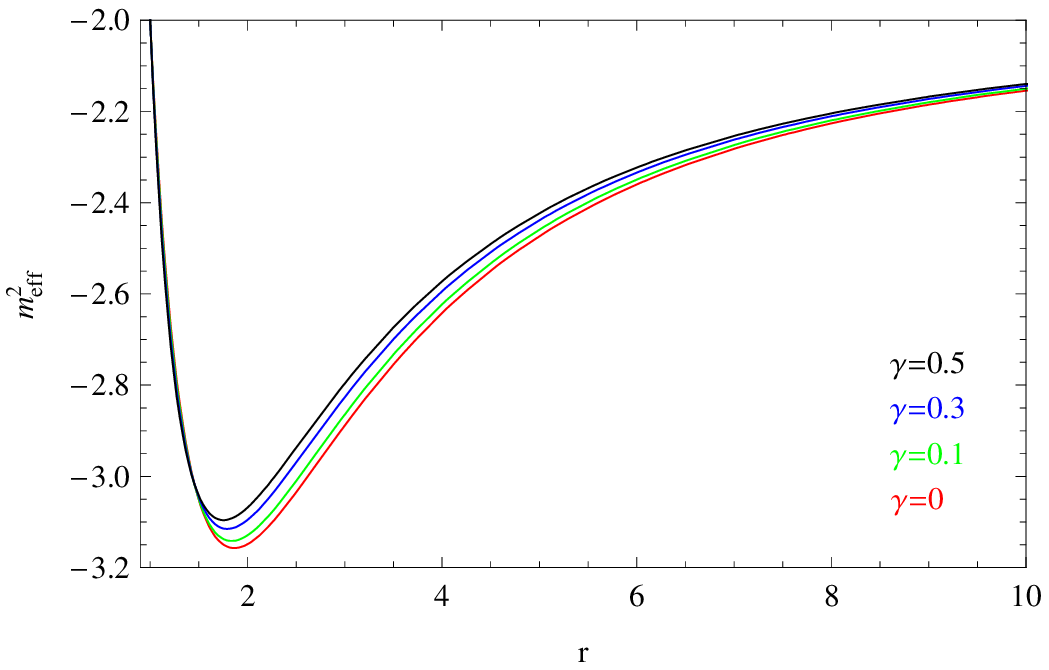}\\ \vspace{0.0cm}
\caption{\label{PsiPhi} (color online) The metric function
$\chi(r)$, the matter functions $\psi(r)$ and $\phi(r)$, the
effective mass $m^{2}_{\rm eff}$ for different values of the
backreacting parameter $\gamma$, i.e., $\gamma=0$ (red), $0.1$
(green), $0.3$ (blue) and $0.5$ (black) with the fixed value of
$\psi_{+}=3.0$.}
\end{figure}

In our computation we fix $r_+=1$. In order to show the effect of
the backreaction, we choose different values of $\gamma$ in
presenting our numerical results. In Fig. \ref{PsiPhi}, we give the
typical solutions with different values of the backreaction
$\gamma=0$ (red), $0.1$ (green), $0.3$ (blue) and $0.5$ (black) for
the metric function $\chi(r)$, the scalar field $\psi(r)$, the
Maxwell field $\phi(r)$ and the effective mass of the scalar field
described by
\begin{eqnarray}
m^{2}_{\rm
eff}=m^{2}+g^{tt}A^{2}_{t}=-\frac{2}{L^{2}}-\frac{\phi^{2}}{ge^{-\chi(r)}}.
\end{eqnarray}
For clarity, we set $\psi_+=3.0$, $c_{4}=0$ and $m^2L^2=-2$ in our
calculation. The solutions for other values of $\psi_{+}$, $c_{4}$
and the scalar mass $m$ are qualitatively similar. Considering the
backreaction, $\chi(r)$ in the metric functions becomes nonzero.
With the increase of the backreaction, we see that $\chi$ deviates
more from zero near the black hole horizon, this indicates that with
stronger backreaction the black hole deviates more from the usual
Schwarzschild AdS black hole as we observed in the probe limit. For
the Maxwell field, $\phi(r)=0$ at the horizon, and we observed that
it increases slower near the horizon for the stronger backreaction.
This shows that stronger backreaction will hinder the growth of the
Maxwell field near the horizon. Although the scalar field $\psi$
will increase with the backreaction near the horizon, the coupling
between the Maxwell field and the scalar field will be reduced when
the backreaction is enhanced, which leads the effective mass to
develop more shallow and narrow wells out of the horizon. The
negative effective mass is the crucial effect to cause the formation
of the scalar hair and the more negative effective mass will make it
easier for the scalar hair to form \cite{GubserPRD78}. The
dependence of the effective mass on the backreaction shows that with
the stronger backreaction the scalar condensate will develop harder.

\begin{figure}[h]
\includegraphics[scale=0.75]{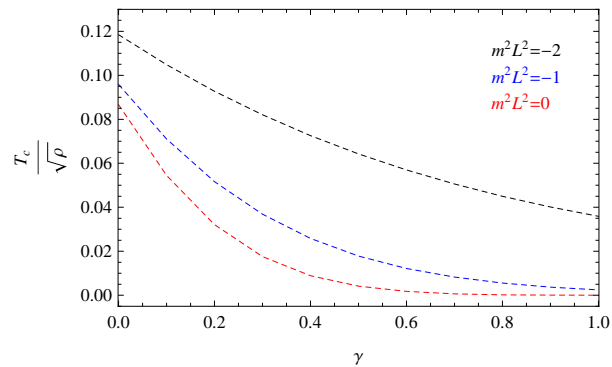}\\ \vspace{0.0cm}
\caption{\label{TcD4} (Color online) The critical temperature
$T_{c}$ as a function of the backreaction $\gamma$ for different
scalar mass $m$. The three dashed lines from top to bottom
correspond to increasing mass, i.e., $m^{2}L^{2}=-2$ (black),
$m^{2}L^{2}=-1$ (blue) and $m^{2}L^{2}=0$ (red), respectively.}
\end{figure}

To see the effect of the backreaction on the scalar condensation
more directly, we plot the behavior of the critical temperature
$T_{c}$ with the change of the backreaction for different scalar
mass $m$ in Fig. \ref{TcD4}. It is clear that for the same scalar
mass $m$, the critical temperature $T_{c}$ drops if $\gamma$
increases, which shows that the backreaction makes the scalar
condensation harder. Fitting the numerical data, we have
\begin{eqnarray}
&&T_{c}\approx0.118\cdot\exp(-1.21\cdot\gamma)\sqrt{\rho},\quad {\rm
for}~~m^{2}L^{2}=-2,\nonumber\\
&&T_{c}\approx0.0974\cdot\exp(-3.33\cdot\gamma)\sqrt{\rho},\quad
{\rm
for}~~m^{2}L^{2}=-1,\nonumber\\
&&T_{c}\approx0.0882\cdot\exp(-5.31\cdot\gamma)\sqrt{\rho},\quad
{\rm for}~~m^{2}L^{2}=0,
\end{eqnarray}
which shows the influence of the backreaction on the critical
temperature $T_{c}$. The exponential dependence of $T_c$ on $\gamma$
is the same as that disclosed in the $3+1$ dimensions
\cite{Brihaye}. For the fixed $\gamma$, $T_{c}$ decreases when
$m^{2}$ becomes less negative, which shows that the larger mass of
the scalar field can make the scalar hair harder to form
\cite{HorowitzPRD78}.

\begin{figure}[H]
\includegraphics[scale=0.75]{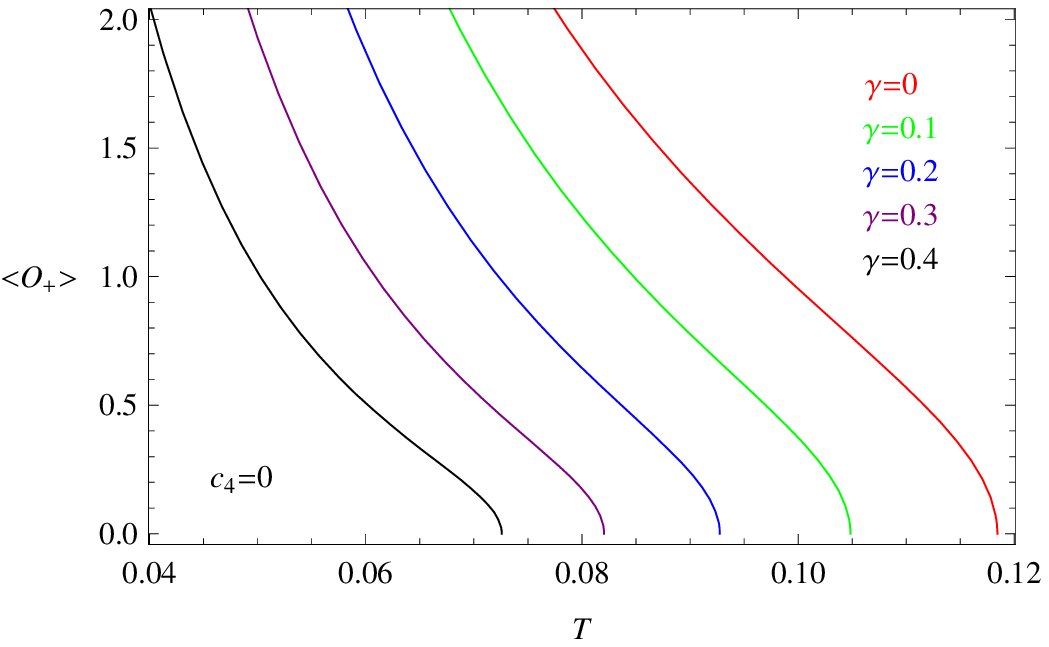}\hspace{0.2cm}%
\includegraphics[scale=0.75]{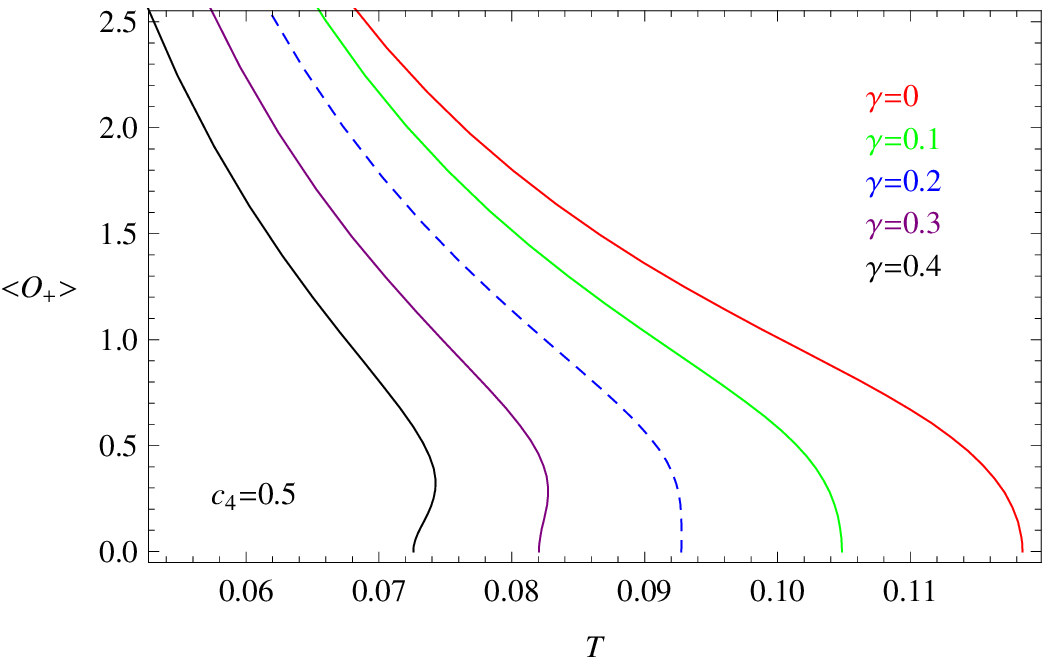}\\ \vspace{0.0cm}
\includegraphics[scale=0.75]{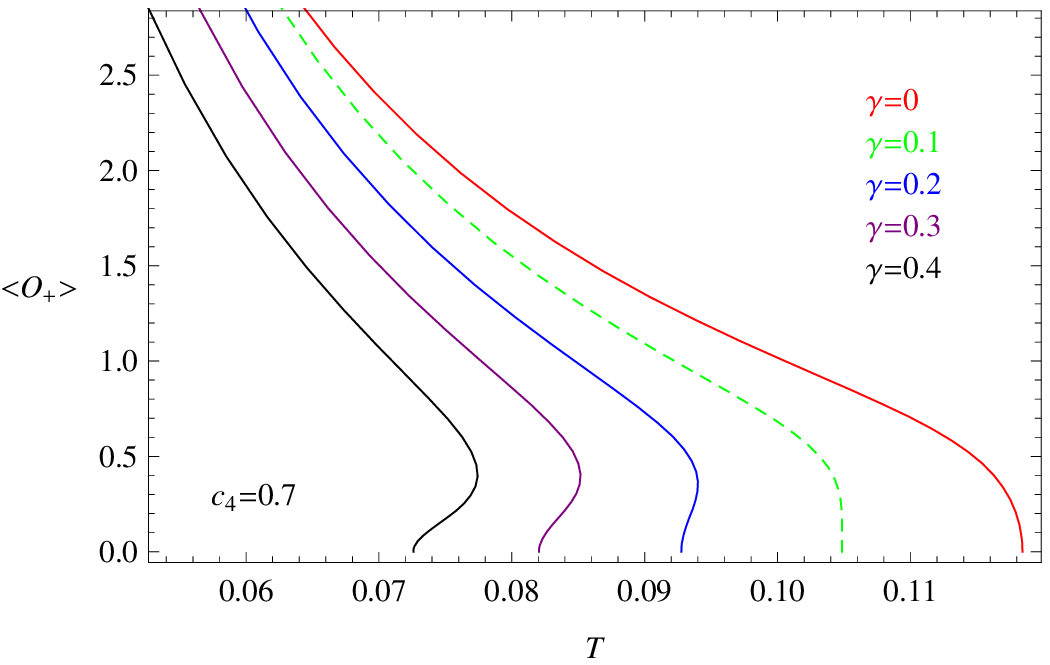}\hspace{0.2cm}%
\includegraphics[scale=0.75]{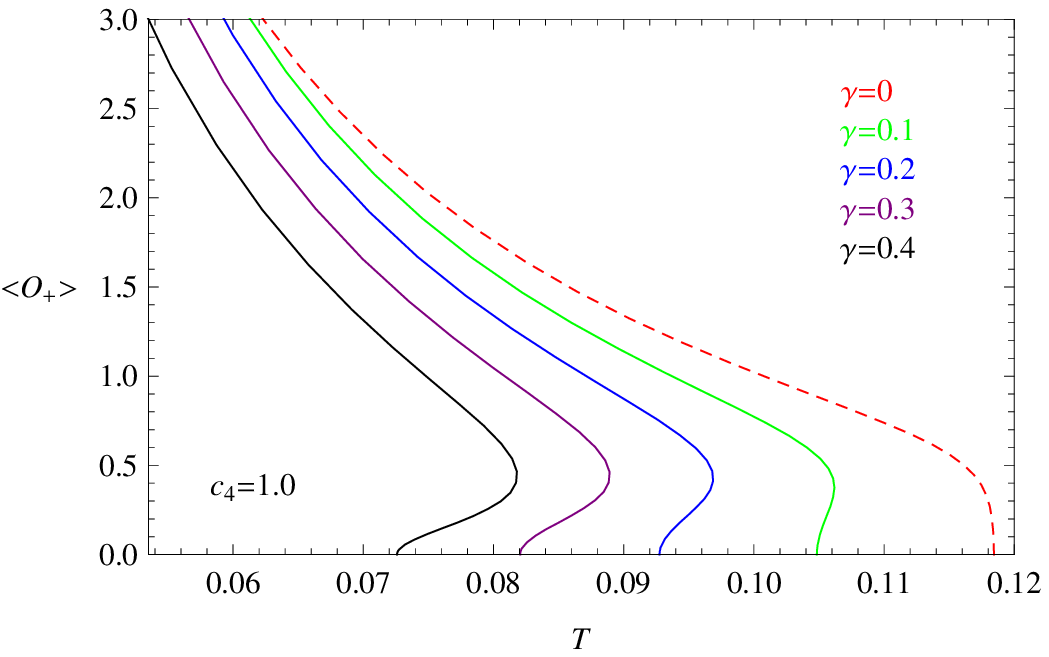}\\ \vspace{0.0cm}
\caption{\label{CondD4} (Color online) The condensate
$<\mathcal{O}_{+}>$ as a function of temperature with fixed values
$c_{4}$ for different values of $\gamma$, which shows that a
different value of $\gamma$ can separate the first- and second-order
behavior. The five lines in each panel from right to left correspond
to increasing $\gamma$, i.e., $0$ (red), $0.1$ (green), $0.2$
(blue), $0.3$ (purple) and $0.4$ (black). For clarity the dashed
line in these panels corresponds to the case of the critical value
$\gamma_{c}$ which can separate the first- and second-order behavior
for different $\mathfrak{F}(\psi)$.}
\end{figure}

It is of great interest to see the influence of the backreaction on the phase transition. In the generalized St\"{u}ckelberg mechanism, broader
descriptions of phase transitions were provided. Here we will discuss whether the backreaction can play the role in the description of the phase
transition. In Fig. \ref{CondD4} we exhibit the condensate of $<\mathcal{O}_{+}>$ for selected values of the parameter $c_{4}$ in
$\mathfrak{F}(\psi)$ and the backreaction $\gamma$. Keeping $c_{4}=0$, we find that the phase transition is always of the second order. The
condensate approaches zero as $<\mathcal{O}_{+}>\propto(T_{c}-T)^{1/2}$ no matter how big is the backreaction we consider. This shows that when
we take $\mathfrak{F}(\psi)=\psi^2$ the backreaction will not change the mean field result allowing the second order phase transition with the
critical exponent $\beta=1/2$ as predicted in the probe limit \cite{HartnollPRL101}. When  the $\psi^4$ term appears in $\mathfrak{F}(\psi)$ with
the strength $c_{4}\geq 1.0$, the condensate $<\mathcal{O}_{+}>$ does not drop to zero continuously at the critical temperature and this behavior
does not alter for choosing different values of $\gamma$. This phenomenon also appears when we consider the condensate $<\mathcal{O}_{-}>$ but
with different $c_{4}$ range. In the probe limit, this phenomenon was attributed to the change of the phase transition from the second order to
the first order \cite{FrancoPRD}. Here we find that the backreaction cannot influence the result when the strength of the $c_{4}\geq 1$.  When
$0<c_{4}<1$, we see that the phenomenon appearing in the condensate to exhibit the change of the second order phase transition to the first order
emerges when the backreaction is strong enough. This observation supports the argument in the study of the holographic p-wave superfluids that
the backreaction plays the role in the phase transition \cite{Ammon-et}. For selected values of $c_{4}$ within the range of $0<c_{4}<1$, we get
the critical value of $\gamma$ to allow the change of the order of the phase transition, i.e., $\gamma_{c}=0.2$ for $c_{4}=0.5$, $\gamma_{c}=0.1$
for $c_{4}=0.7$ and $\gamma_{c}=0$ for $c_{4}=1$. It shows that $\gamma_{c}$ becomes smaller when $c_{4}$ is bigger. Thus, when the strength of
the $\psi^4$ term is not strong enough in the $\mathfrak{F}(\psi)$, the backreaction will combine with the strength $c_{4}$ to tune the order of
the phase transition. With the backreaction of the spacetime, we see richer physics in the phase transition.

\subsection{The effects on the conductivity}

In order to investigate the influence of the backreaction on the
conductivity, we consider the time-dependent perturbation with zero
momentum $A_{x}=a_{x}(r)e^{-i\omega t}$ and $g_{tx}=f(r)e^{-i\omega
t}$ which can get the equations of motion decoupled from other
perturbations \cite{Aprile-Russo,Liu-Sun,HartnollJHEP12}
\begin{eqnarray}
a_{x}^{\prime\prime}+\left(\frac{g^\prime}{g}-\frac{\chi^\prime}{2}\right)a_{x}^\prime
+\left[\frac{\omega^2}{g^2}e^{\chi}-\frac{\mathfrak{F}(\psi)}{g}\right]a_{x}
+\frac{\phi^\prime}{g}e^{\chi}\left(f^\prime-\frac{2f}{r}\right)=0,
\label{Maxwell Equation}
\end{eqnarray}
\begin{eqnarray}
f^\prime-\frac{2f}{r}+\gamma\phi^\prime a_{x}=0. \label{Metric
perturbation}
\end{eqnarray}
Substituting Eq. (\ref{Metric perturbation}) into Eq. (\ref{Maxwell
Equation}), we have the  equation of motion for the perturbed
Maxwell field
\begin{eqnarray}
a_{x}^{\prime\prime}+\left(\frac{g^\prime}{g}-\frac{\chi^\prime}{2}\right)a_{x}^\prime
+\left[\left(\frac{\omega^2}{g^2}-\frac{\gamma\phi^{\prime
2}}{g}\right)e^{\chi}-\frac{\mathfrak{F}(\psi)}{g}\right]a_{x}=0.
\label{Revised Maxwell Equation}
\end{eqnarray}
Near the horizon, we solve the above equation by imposing the
ingoing boundary condition
\begin{eqnarray}
a_{x}(r)\propto g(r)^{-i\omega/(4\pi T_{H})}.
\end{eqnarray}
In the asymptotic AdS region, the asymptotic behavior of the
perturbations can be expressed as
\begin{eqnarray}
a_{x}=a^{(0)}_{x}+\frac{a^{(1)}_{x}}{r},~~f=r^{2}f^{(0)}+\frac{f^{(1)}}{r}.
\end{eqnarray}
Thus, we have the conductivity of the dual superconductor by using
the AdS/CFT dictionary \cite{HartnollJHEP12}
\begin{eqnarray}\label{Conductivity}
\sigma(\omega)=-\frac{ia^{(1)}_{x}}{\omega a^{(0)}_{x}}.
\end{eqnarray}
For the general forms of the function
$\mathfrak{F}(\psi)=\psi^{2}+c_{4}\psi^{4}$, we can obtain the
conductivity by solving the Maxwell equation numerically. In our
computation we fix $m^{2}L^{2}=-2$. Other values of the scalar mass
present the same qualitative results \cite{HorowitzPRD78,Pan-Wang}.

\begin{figure}[H]
\includegraphics[scale=0.5]{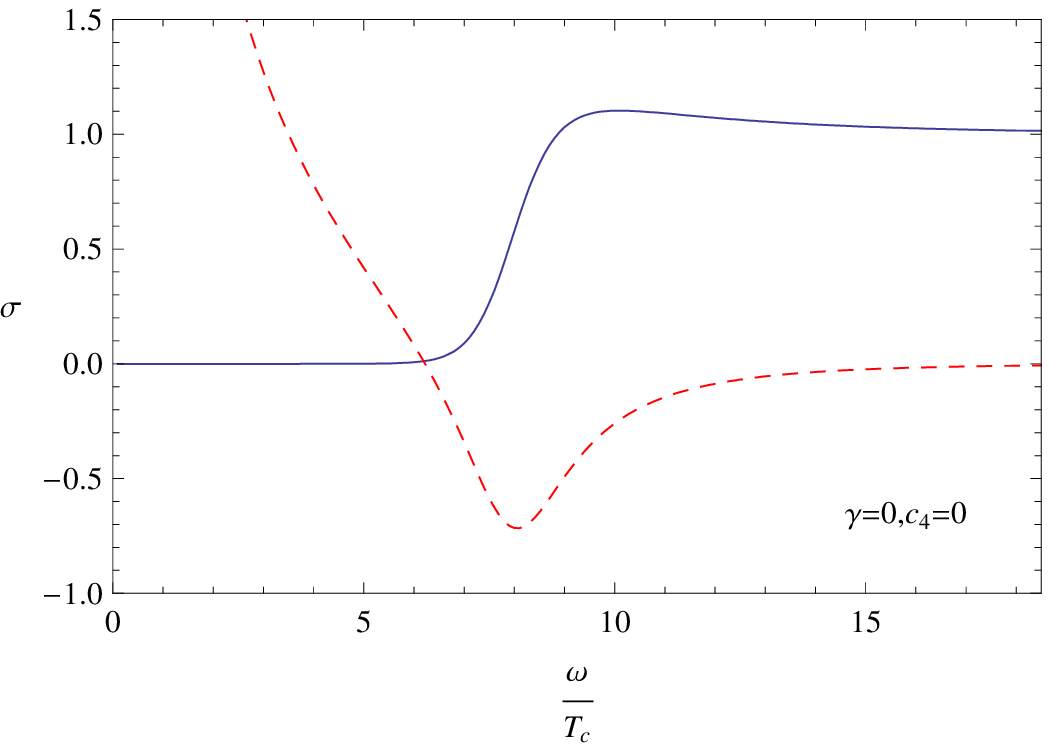}\hspace{0.2cm}%
\includegraphics[scale=0.5]{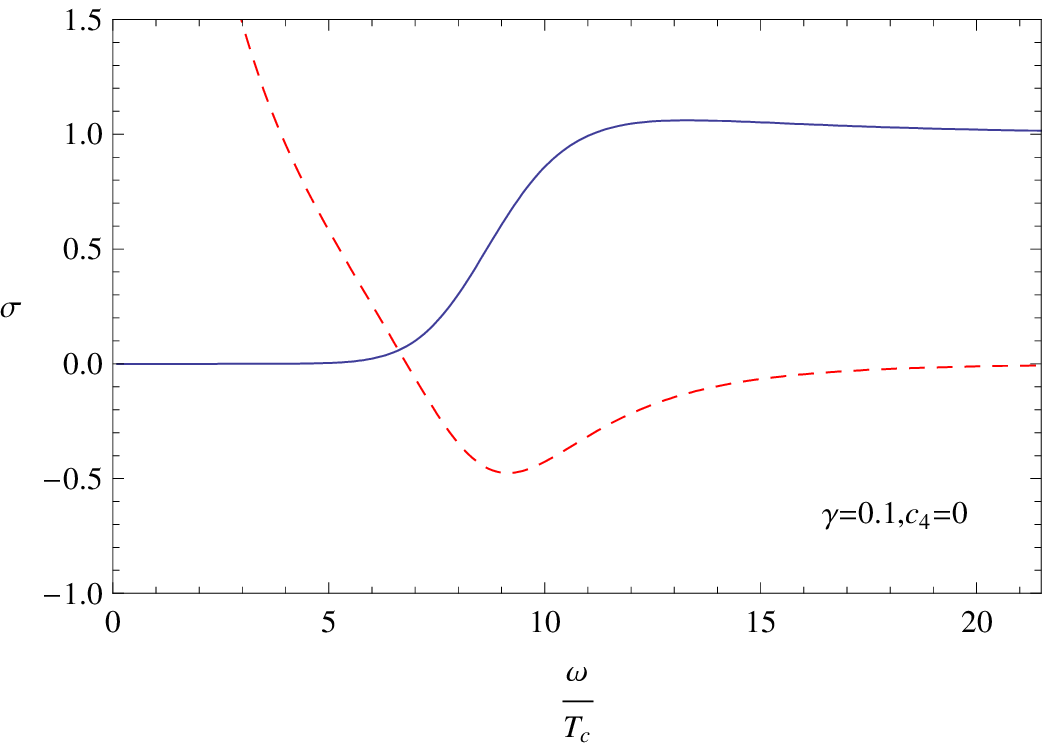}\hspace{0.2cm}%
\includegraphics[scale=0.5]{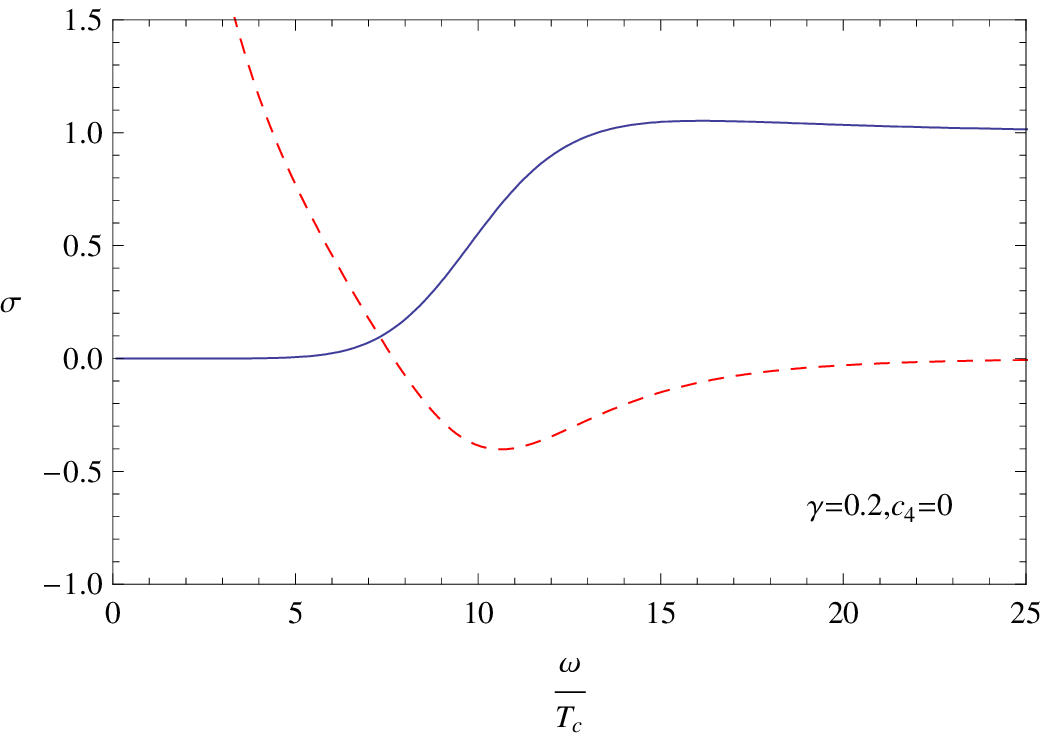}\\ \vspace{0.0cm}
\includegraphics[scale=0.5]{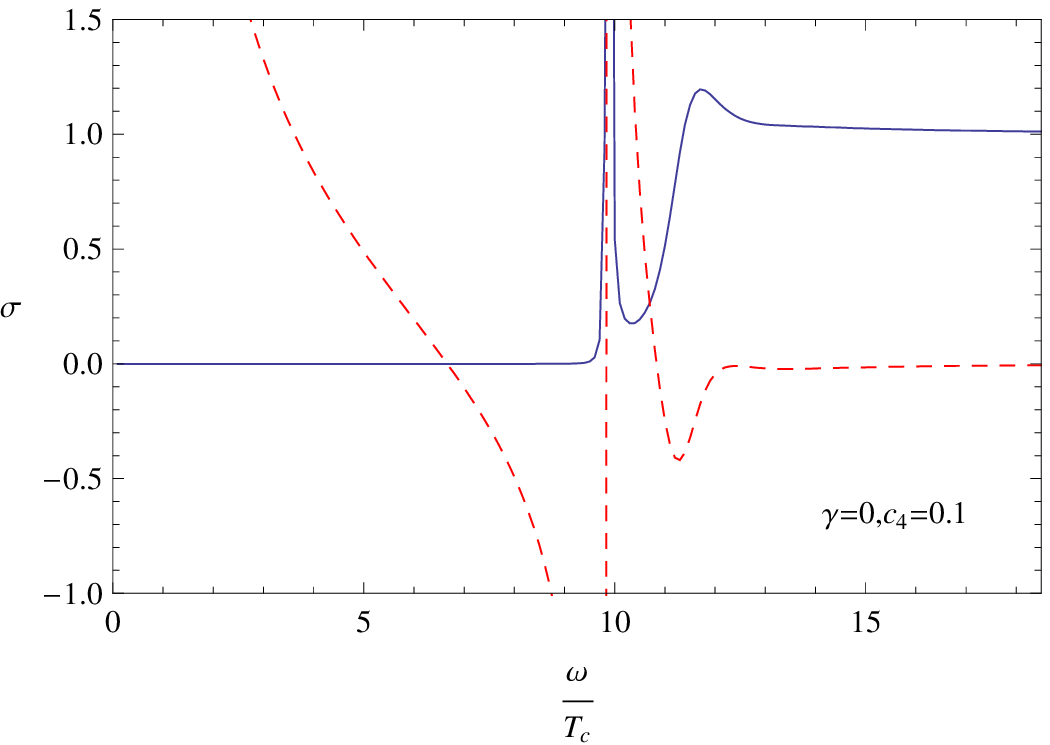}\hspace{0.2cm}%
\includegraphics[scale=0.5]{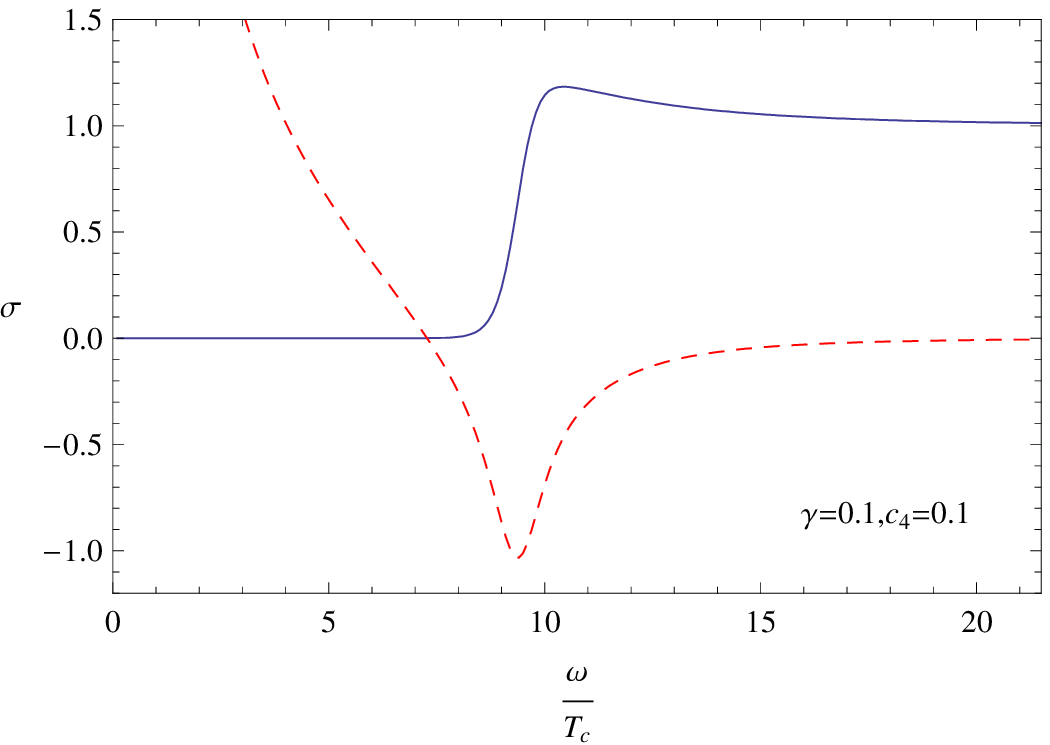}\hspace{0.2cm}%
\includegraphics[scale=0.5]{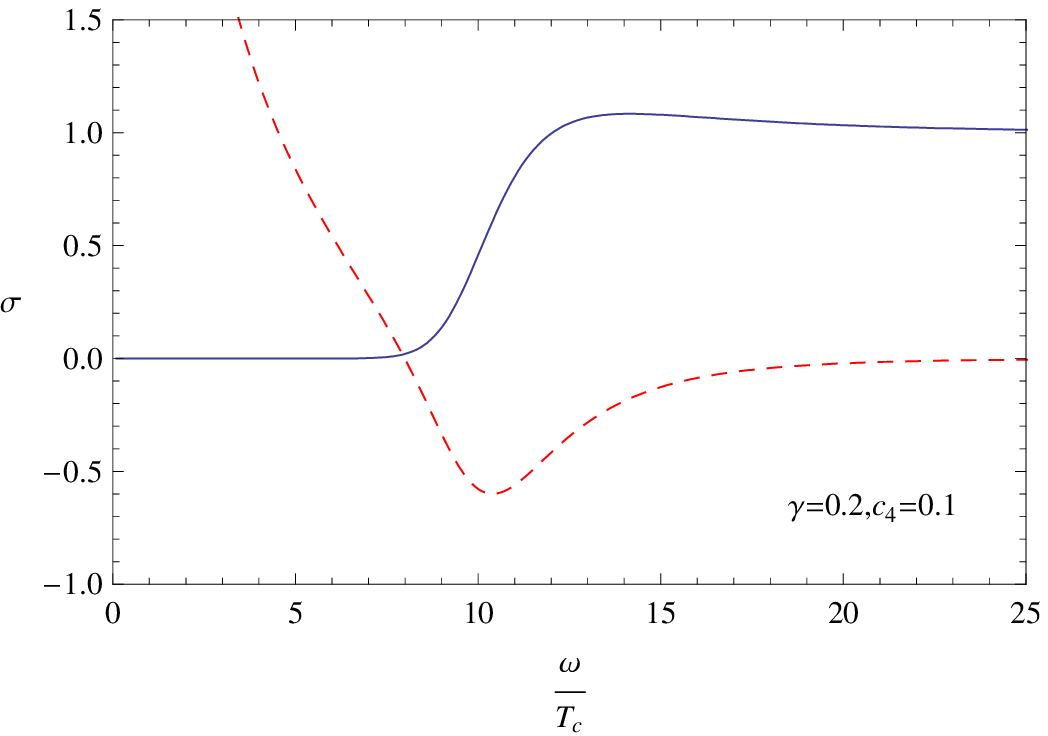}\\ \vspace{0.0cm}
\caption{\label{c4-Gamma-Conductivity} (color online) Conductivity
for ($2+1$)-dimensional superconductors with fixed values of the
model parameter $c_{4}$ for different backreaction $\gamma$. The
blue (solid) line and red (dashed) line represent the real part and
imaginary part of the conductivity respectively.}
\end{figure}

In the probe limit, it was argued in \cite{HorowitzPRD78} that there
is a universal relation between the gap $\omega_g$ in the frequency
dependent conductivity and the critical temperature $T_c$,
$\omega_g/T_{c} \approx 8$, which is roughly two times bigger than
the BCS value $3.5$ indicating that the holographic superconductors
are strongly coupled. However this so-called universal relation was
challenged when the higher curvature corrections are taken into
account \cite{Pan-Wang,Gregory}. When taking the backreaction of the
spacetime into account, we find that the claimed universal relation
can be modified even without the high curvature correction. In Fig.
\ref{c4-Gamma-Conductivity} we plot the frequency dependent
conductivity obtained by solving the Maxwell equation numerically
for $c_{4}=0$ and $0.1$ with different strength of the backreaction,
i.e., $\gamma=0$, $0.1$ and $0.2$ respectively at temperature
$T/T_{c}\simeq0.2$ . It clearly shows that for fixed $c_{4}$,  the
gap frequency $\omega_g$ becomes larger when the backreaction is
stronger. The deviation from $\omega_g/T_{c}\approx8$ becomes bigger
with the increase of $\gamma$. This behavior is consistent with the
observation in five dimensional Einstein-Gauss-Bonnet gravity
\cite{Barclay-Gregory}. Here we show that the backreaction also
changes the ratio $\omega_g/T_{c}$ in lower dimensional backgrounds
without higher curvature corrections. For the selected $\gamma$, the
ratio $\omega_g/T_{c}$ also depends on the model parameter $c_{4}$,
which agrees with the result obtained for general holographic
superconductor models with Gauss-Bonnet corrections in the probe
limit \cite{Pan-WangPLB}.

So far, we conclude that the gap ratio $\omega_g/T_{c}$ does not
only change in the AdS spacetimes with higher curvature correction,
it also alters in the presence of the backreaction and the $\psi^4$
term in the $\mathfrak{F}(\psi)$. Thus there is no universal
relation $\omega_g/T_{c} \approx 8$ for general holographic
superconductors.

\section{Gauss-Bonnet superconducting models with backreaction}

In the following we generalize the above discussion to the
Gauss-Bonnet superconducting models. We extend the Lagrangian in Eq.
(\ref{System}) to ($4+1$)-dimensional Einstein-Gauss-Bonnet gravity
\begin{eqnarray}\label{SystemGB}
S=\frac{1}{16\pi G_{5}}\int
d^{5}x\sqrt{-g}\left[R+\frac{12}{L^{2}}+\frac{\alpha}{2}\left(R_{\mu\nu\lambda\delta}R^{\mu\nu\lambda\delta}
-4R_{\mu\nu}R^{\mu\nu}+R^{2}\right)+16\pi G_{5}\mathcal{L}_{matter}
\right],
\end{eqnarray}
where $G_{5}$ is the 5-dimensional Newton constant, $\alpha$ is the
Gauss-Bonnet coupling constant and $\mathcal{L}_{matter}$ is the
same generalized St\"{u}ckelberg Lagrangian given in Eq.
(\ref{ModelL}). The Gauss-Bonnet parameter $\alpha$ has an upper
bound called the Chern-Simons limit $\alpha=L^{2}/4$ which can
guarantee a well-defined vacuum for the gravity theory
\cite{Cai-2002}, and a lower bound $\alpha=-7L^{2}/36$ determined by
the causality \cite{BrigantePR,Buchel-Myers,Camanho-Edelstein,
Buchel-Escobedo-Myers,Hofman,Boer-Kulaxizi}.

Taking the Ansatz for the metric in the five-dimensional spacetime
\begin{eqnarray}\label{GB-BH metric}
ds^2=-g(r)e^{-\chi(r)}dt^{2}+\frac{dr^2}{g(r)}+r^{2}(dx^{2}+dy^{2}+dz^{2}),
\end{eqnarray}
we can have the Hawking temperature as expressed in Eq. (\ref{Hawking temperature}) and get the equations of motion
\begin{eqnarray}\label{GBPsi-Phi}
&&\chi^{\prime}+\frac{2}{3}\gamma\frac{r^{2}}{r^{2}-2\alpha
g}\left[\frac{r}{2}\psi^{\prime
2}+\frac{r}{2g^{2}}e^{\chi}\phi^{2}\mathfrak{F}(\psi)\right]=0,\nonumber\\
&&g^{\prime}-\frac{r^{2}}{r^{2}-2\alpha
g}\left\{\left(\frac{4r}{L^{2}}-\frac{2g}{r}\right)-\frac{2}{3}\gamma
rg\left[\frac{1}{4}\psi^{\prime 2}+\frac{1}{4g}e^{\chi}\phi^{\prime
2}+\frac{m^{2}}{4g}\psi^{2}+\frac{1}{4g^{2}}e^{\chi}\phi^{2}\mathfrak{F}(\psi)\right]\right\}=0,
\nonumber\\ &&
\phi^{\prime\prime}+\left(\frac{3}{r}+\frac{\chi^{\prime}}{2}\right)\phi^\prime-\frac{\mathfrak{F}(\psi)}{g}\phi=0,\nonumber\\
&& \psi^{\prime\prime}+\left(\frac{3}{r}-\frac{\chi^{\prime}}{2}+
\frac{g^\prime}{g}\right)\psi^\prime
-\frac{m^2}{g}\psi+\frac{1}{2g^2}e^{\chi}\phi^2\mathfrak{F}^{\prime}(\psi)=0,
\end{eqnarray}
where we have set the backreaction $\gamma=16\pi G_{5}$. In the
limit $\gamma\rightarrow0$, (\ref{GBPsi-Phi}) reduce to describe the
general ($3+1$)-dimensional holographic superconductor model with
Gauss-Bonnet corrections in the probe limit \cite{Pan-WangPLB}.

An exact solution to Eq. (\ref{GBPsi-Phi}) is the charged Gauss-Bonnet black hole described by \cite{Brihaye}
\begin{eqnarray}
g=\frac{r^{2}}{2\alpha}\left(1-\sqrt{1-\frac{4\alpha}{L^{2}}+\frac{4\alpha
M}{r^{4}}-\frac{4\gamma\alpha\rho^{2}}{3r^{6}}}\right)\,,\hspace{0.5cm}
\phi=\rho\left(\frac{1}{r^{2}_{+}}-\frac{1}{r^{2}}\right)\,,\hspace{0.5cm}\chi=\psi=0.
\end{eqnarray}
Obviously, it is not easy to find other analytic solutions to these
nonlinear equations. So we have to count on the numerical
computation. The boundary conditions at the asymptotic AdS boundary
($r\rightarrow\infty$) are
\begin{eqnarray}
\psi=\frac{\psi_{-}}{r^{\lambda_{-}}}+\frac{\psi_{+}}{r^{\lambda_{+}}}\,,\hspace{0.5cm}
\phi=\mu-\frac{\rho}{r^{2}}\,, \label{GBinfinity}
\end{eqnarray}
with
\begin{eqnarray}
\lambda_{\pm}=2\pm\sqrt{4+m^{2}L^2_{\rm eff}},
\end{eqnarray}
where $L_{\rm eff}^{2}=2\alpha/(1-\sqrt{1-4\alpha/L^2}~)$ is the
effective asymptotic AdS scale \cite{Cai-2002}. We will concentrate
on the scalar condensate $\langle{\cal O}_{+}\rangle=\psi_{+}$ and
set $\psi_{-}=0$. For concreteness, we will take the fixed mass of
the scalar field $m^2L^2=-3$. The alternative choice of the mass
$m^2L_{\rm eff}^2=-3$ will not qualitatively change our results
\cite{Gregory,Pan-Wang}.

\begin{figure}[H]
\includegraphics[scale=0.52]{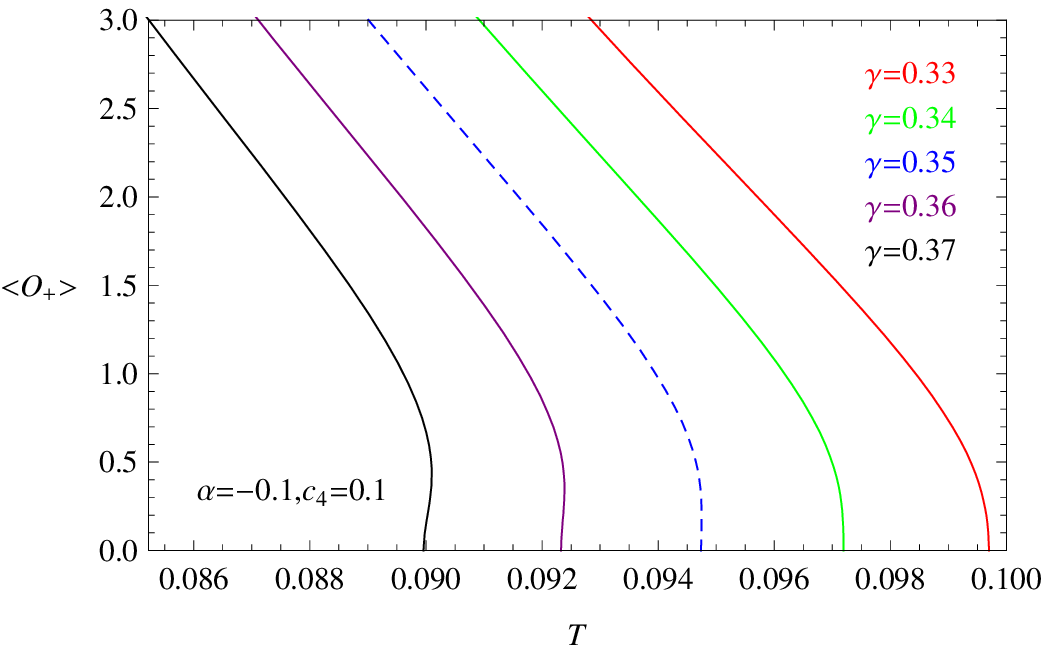}\hspace{0.2cm}%
\includegraphics[scale=0.52]{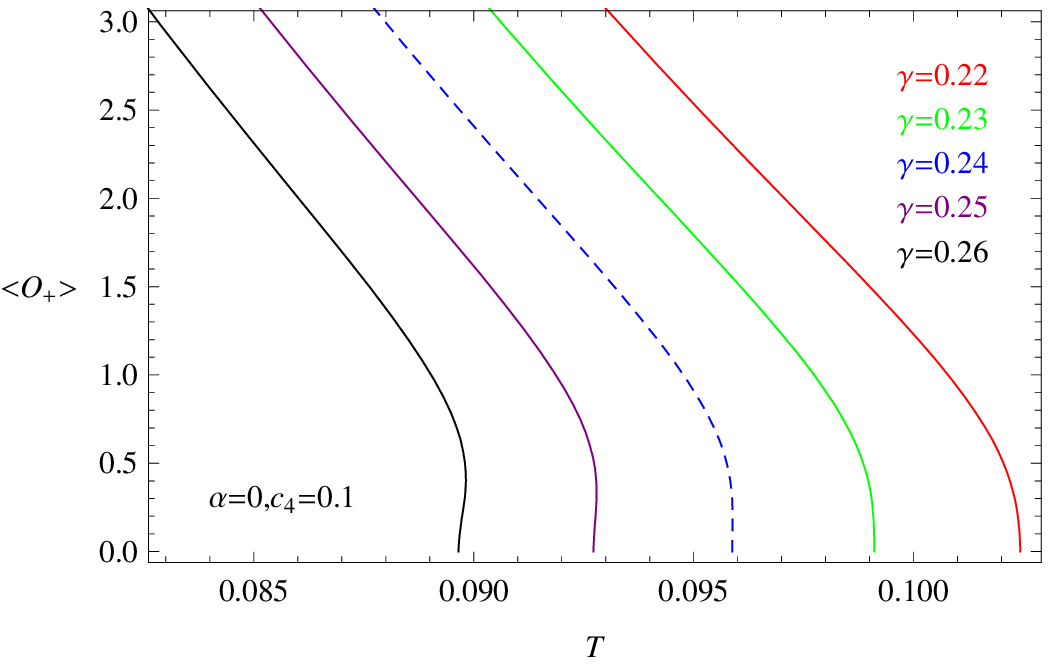}\vspace{0.2cm}%
\includegraphics[scale=0.52]{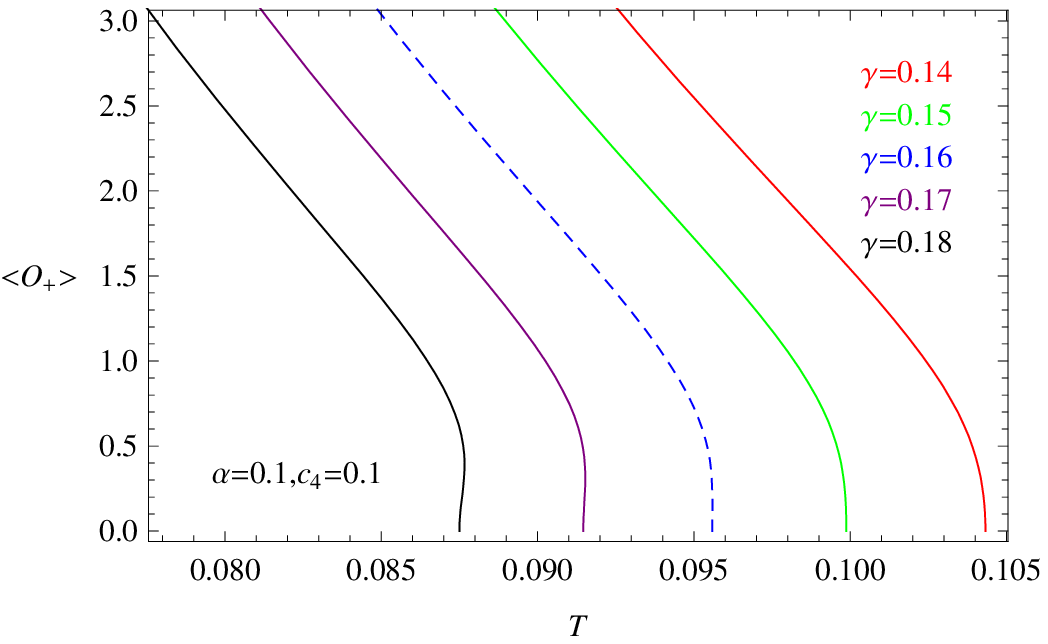}\\ \vspace{0.0cm}
\includegraphics[scale=0.52]{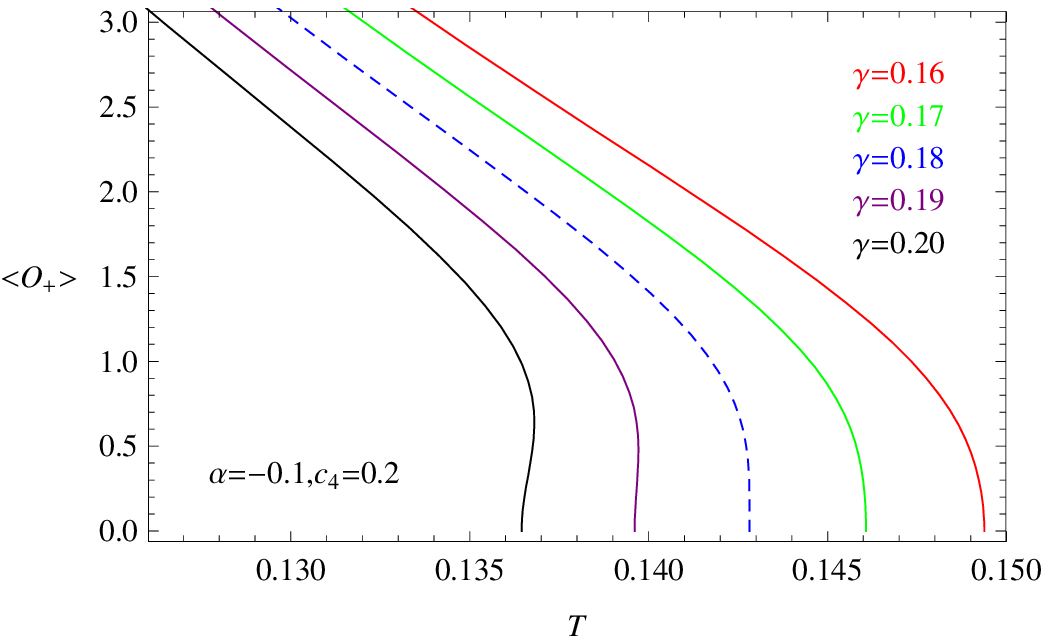}\hspace{0.2cm}%
\includegraphics[scale=0.52]{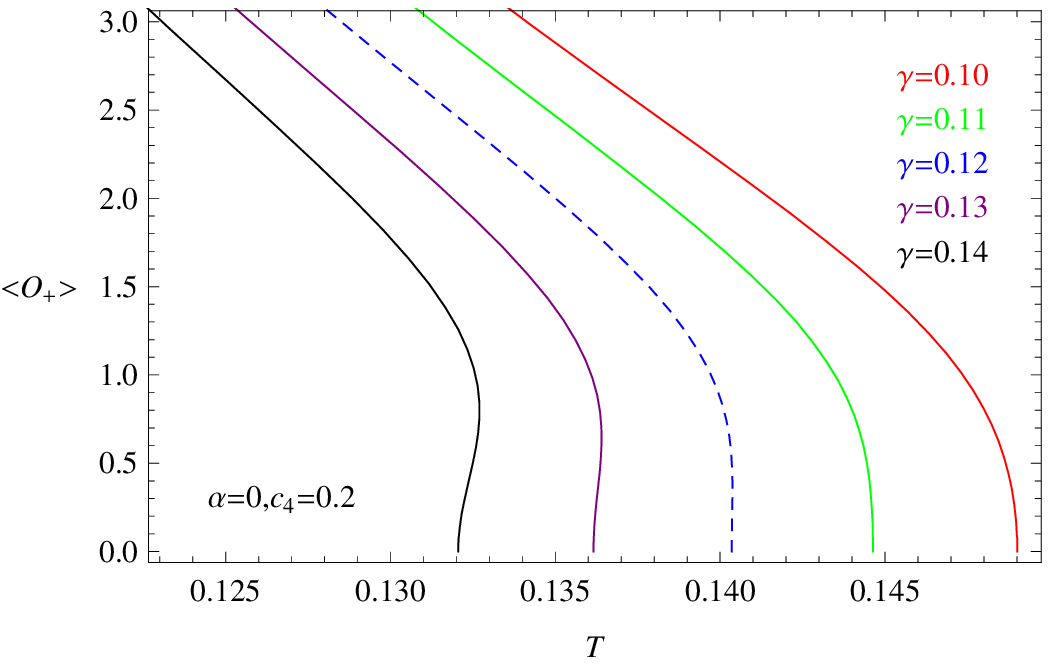}\vspace{0.2cm}%
\includegraphics[scale=0.52]{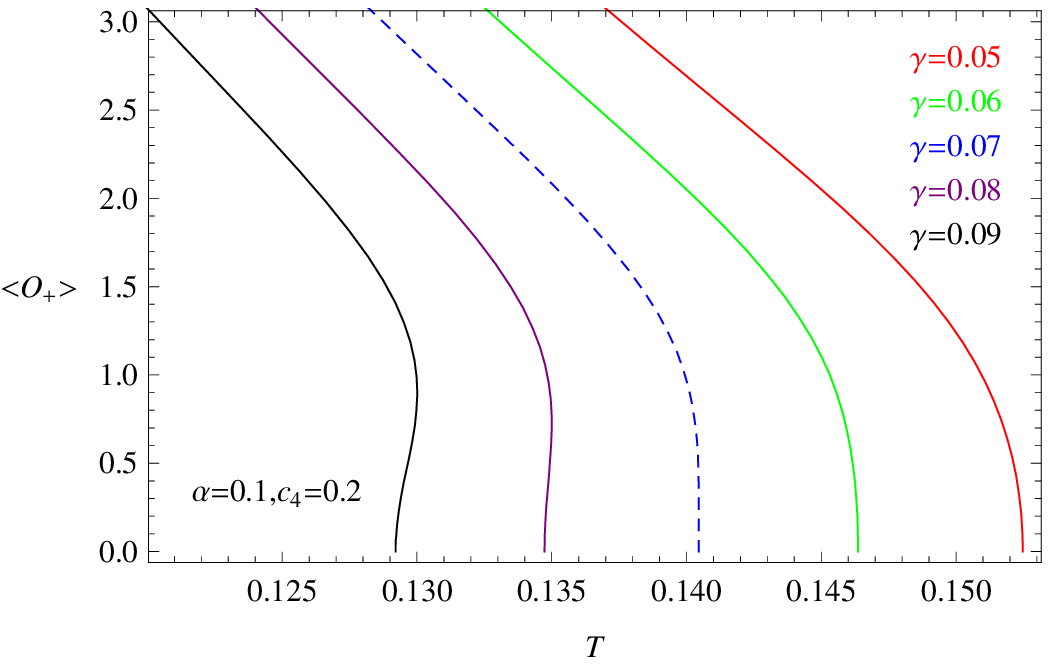}\\ \vspace{0.0cm}
\caption{\label{CondGBcLambda} (Color online) The condensate
$<\mathcal{O}_{+}>$ as a function of temperature with fixed
Gauss-Bonnet correction term $\alpha$ and model parameter $c_{4}$
for different values of $\gamma$, which shows that the critical
value $\gamma_{c}$ (the blue and dashed line in each panel)
decreases as $\alpha$ or $c_{4}$ increases. The five lines in each
panel from right to left correspond to increasing $\gamma$ for fixed
$\alpha$ and $c_{4}$.}
\end{figure}

\begin{table}[ht]
\caption{\label{critical value} The critical value $\gamma_{c}$
which can separate the first- and second-order behavior for
different Gauss-Bonnet correction term $\alpha$ with fixed
$\mathfrak{F}(\psi)=\psi^{2}+c_{4}\psi^{4}$. }
\begin{tabular}{|c|c|c|c|c|c|}
         \hline
$~~\alpha~~$~~&~~-0.1~~&~~0~~&~~0.1~~&~~0.2~~&~~0.25~~
          \\
        \hline
~~$c_{4}=0.1$~~&~~0.35~~&~~0.24~~&~~0.16~~&~~0.09~~&~~0.05~~
          \\
        \hline
~~$c_{4}=0.2$~~&~~0.18~~&~~0.12~~&~~0.07~~&~~0.03~~&~~0.01~~
          \\
        \hline
\end{tabular}
\end{table}

In the probe limit we have already observed that different values of
Gauss-Bonnet correction term and model parameters can determine the
order of phase transition \cite{Pan-WangPLB}. Considering the
backreaction, we obtain richer descriptions in the phase transition.
When $c_4=0$, no matter what values of the Gauss-Bonnet factor and
the backreaction we choose, $<\mathcal{O}_{+}>$ always drops to zero
continuously at the critical temperature, which indicates that the
phase transition is always of the second order. When $c_4$ deviates
from zero, for chosen $\alpha$, $<\mathcal{O}_{+}>$ can become
multivalued near the critical temperature when the backreaction is
strong enough, which indicates that the first order of the phase
transition can happen. In Fig. \ref{CondGBcLambda}, we exhibit the
results of the condensate $<\mathcal{O}_{+}>$ for chosen values of
$c_4$ and $\alpha$ but variable strength of the backreaction
$\gamma$. The critical values of the backreaction to allow the
condensate not dropping to zero continuously at the critical
temperature are listed in Table \ref{critical value} for different
values of Gauss-Bonnet correction term $\alpha$ with the selected
$c_{4}=0.1$ and $0.2$. For bigger values of  $\alpha$ or $c_4$, we
observe that the critical $\gamma$ is smaller to accommodate the
first order phase transition.

In the Gauss-Bonnet gravity, we conclude that the backreaction
$\gamma$ together with the Gauss-Bonnet factor $\alpha$ and the
model parameter $c_4$ plays the role in determining the order of the
phase transition.

\section{conclusions}

In this work we have studied the general holographic superconductors away from the probe limit. We have considered the four-dimensional and
five-dimensional Einstein and Einstein-Gauss-Bonnet gravity backgrounds. We observed that the backreaction can make the condensation harder to be
formed. In addition to the model parameters in $\mathfrak{F}$ and the Gauss-Bonnet factor, we found that the spacetime backreaction can also
bring richer descriptions in the phase transition. When the curvature correction term or model parameter is larger, smaller backreaction can
trigger the first order phase transition.  This observation supports the finding in the holographic p-wave superfluids with backreactions in
$3+1$ dimensions \cite{Ammon-et}. Extending the analysis to the conductivity, we further disclosed the fact that there is no universal relation
for $\omega_g/T_{c}$ in the holographic superconductor when the backreaction is taken into account.

\begin{acknowledgments}
We acknowledge the KITPC/ITP, Beijing, for its great hospitality and
stimulating scientific environment where the work was finished. This
work was partially supported by the National Natural Science
Foundation of China under Grant Nos. 11075036 and 10905020. Qiyuan
Pan was also supported by the China Postdoctoral Science Foundation
under Grant No. 201003245.

\end{acknowledgments}

\end{document}